\documentclass[prl,twocolumn,showpacs,preprintnumbers,amsmath,amssymb]{revtex4}
\usepackage{dcolumn}
\usepackage{bm}
\usepackage{graphicx}
\begin{document}

\title{Spin polarization of electron current on the quantum well with exchange-splitted levels}

\author{L.V. Lutsev}
 \email{l_lutsev@mail.ru}

\affiliation{A.F. Ioffe Physical-Technical Institute of the Russian
Academy of Sciences, 194021 St. Petersburg, Russia \\}

\date{\today}

\begin{abstract}
The backscattering process of injected electrons on
exchange-splitted levels of quantum well (QW) in ferromagnetic metal
/ insulator / semiconductor heterostructure is studied. It is found
that, if one of the exchange-splitted levels lies in the top region
of the QW and the energy of injected electrons is close to the
energy of localized electron on this level, the backward scattering
becomes dependent on spins of injected electrons. Accumulation of
backscattered electrons in the QW leads to considerable reduction of
the current depended on its spin orientation. The spin polarization
increases with growth of the applied electric field and the storage
time of electrons in the QW. High values of the spin polarization
can be achieved at room temperature. In this way, the QW with
exchange-splitted levels in ferromagnetic metal / insulator /
semiconductor heterostructure can be used as effective spin filter.
\end{abstract}

\pacs{72.25.DC}
\keywords{quantum well, spin-dependent backscattering, exchange-splitted levels, spin polarization}
\maketitle

\section{Introduction}
The active manipulation of spin-dependent electron transport is the
principal task in spintronics~\cite{ref1,ref2,ref3}. One of the
effective way to achieve spin polarization of electrons injected
into semiconductors is spin-dependent tunnelling through a barrier
in ferromagnetic metal / insulator / semiconductor heterostructures
\cite{Hammar,Jiang,Li05,Dong,Jonker,Li09,Kios09,Saito10}. The
maximum of the spin injection efficiency reaches 52\% at 100 K and
32\% at 290 K for a MgO barrier on GaAs~\cite{Jiang}. High
electrical injection of spin-polarized electrons from a Fe film
through an Al${ }_2$O${ }_3$ tunnel barrier into Si has been
demonstrated in~\cite{Jonker}. However, in Si the electron spin
polarization was observed at low temperatures -- 30\% at 5 K, with
polarization extending to at least 125 K.

Although important results in the spin injection have been obtained,
high values of the spin polarization of injected electrons at room
temperature has not been achieved. Therefore, it is crucial to find
a new method of the spin polarization that allows us to achieve high
spin-injection efficiency. In this paper, we present new method of
the electron spin polarization in ferromagnetic metal / insulator /
semiconductor heterostructure, which based on spin-dependent
backward scattering on exchange-splitted levels in a $2D$ quantum
well (QW) and on the electron capture by this QW. The QW is formed
in the semiconductor near the insulator / semiconductor interface.
The insulator layer is thin in order to provide electron tunneling
from the ferromagnetic metal and to split levels in the QW by the
exchange interaction. For the case, when one of exchange-splitted
levels lies in the top region of the QW, the backscattering process
of injected electrons becomes dependent on their spins. The capture
of backscattered spin-polarized electrons by the QW leads to an
additional Coulomb repulsion for electrons tunneling from the
ferromagnetic metal and to a considerable spin-polarized decrease of
the current flowing in the heterostructure. In this way, high values
of the spin polarization and manipulation of injected electron spins
by the charge of the QW can be achieved.

By analogy with the spin polarization in ferromagnetic metal /
insulator / semiconductor heterostructures, spin polarization of
electron current can be observed in heterostructures consisted of
semiconductor substrates and granular films with ferromagnetic metal
nanoparticles in an insulator matrix. Ones of these heterostructures
are SiO${ }_2$(Co)/GaAs heterostructures, where the SiO${ }_2$(Co)
is the granular SiO${ }_2$ film with Co
nanoparticles~\cite{Lut05,Lut06,Lut09,Lut13}. The $2D$ QW
(accumulation electron layer) with exchange-splitted levels is
formed at the interface in the GaAs~\cite{Lut09,Lut13,Lut06a}. In
SiO${ }_2$(Co)/GaAs heterostructures extremely large
magnetoresistance and the current reduction on the temperature
dependence are observed at room temperature.

The paper is organized as follows. In the next section, we study the
electron backscattering process on exchange-splitted levels in $2D$
QW. In Sec. III we consider the capture of backscattered electrons
by the QW and the current reduction caused by the QW charging. The
spin polarization of electron current caused by the backscattering
process and, consequently, by the current reduction is described in
Sec. IV. Finally, we summarize the results in Sec. V.

\section{Influence of localized levels in the quantum well on the electron
 backscattering process}

\begin{figure}
\begin{center}
\includegraphics*[scale=.95]{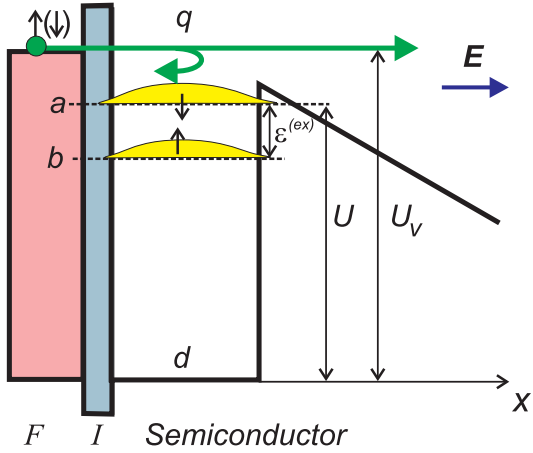}
\end{center}
\caption{ Backward scattering of injected electron on
exchange-splitted levels in the quantum well with the width $d$. $F$
is the ferromagnetic metal, $I$ is the insulator layer, $E$ is the
applied electric field.} \label{bar2}
\end{figure}

Let us consider the backward scattering of injected electrons on
exchange-splitted levels of the QW in a ferromagnetic metal /
insulator / semiconductor heterostructure (Fig. \ref{bar2}).
Exchange interaction between electrons in the ferromagnetic metal
and electrons in the QW through the thin insulator layer splits
electron levels in the QW. Electrons on exchange-splitted sublevels
(sublevels $a$ and $b$ in Fig. \ref{bar2}) have opposite spin
orientations. Difference of their energies is equal to the exchange
energy $\varepsilon^{(ex)}$. For clarification of the main features
of scattering dependencies we restrict our consideration on the
backscattering process on one of exchange-splitted levels -- the top
sublevel $a$ with certain spin orientation and neglect insignificant
parts of the electron wavefunction on the sublevel $a$ outside the
QW.

The electron wavefunction on the sublevel $a$ is the product of the
spatial function $u(x)$ and the spin function $\chi_u(\sigma_u)$

$$\psi(x,\sigma_u)= u(x)\chi_u(\sigma_u),$$

\noindent where $\sigma_u=\uparrow,\downarrow$ is the electron spin.
In the WKB (Wentzel-Kramers-Brillouin) approximation \cite{Dav} the
spatial function in the QW can be written as

\begin{equation}
u(x)=\frac{C_u}{\sqrt{|k_a|}}\sin(k_ax+\pi/4), \label{1bar}
\end{equation}

\noindent where $k_a=\sqrt{2mU}/\hbar=\pi(n+1/2)/d$ is the
wavevector of the electron on the sublevel $a$ in the zero
approximation with respect to $\varepsilon^{(ex)}/U\ll 1$, $m$ is
the electron mass, $U$ is the energy counted from the QW bottom, $d$
is the width of the QW, $C_u$ is the normalization coefficient,
$n=0,1,2,\ldots$ is the number of the sublevel $a$.

The wavefunction of injected electron flying over the QW has the
form of the product of the spatial function $v(x)$ and the spin
function $\chi_v(\sigma_v)$

$$\varphi(x,\sigma_v)=v(x)\chi_v(\sigma_v),$$

\noindent where
\begin{equation}
v(x)=\frac{C_v}{\sqrt{|q|}}\exp(iqx), \label{2bar}
\end{equation}

\noindent $q=\sqrt{2mU_v}/\hbar$, $U_v$ is the energy counted from
the QW bottom, $C_v$ is the normalization coefficient.

For the interaction $W(x)$ between the injected electron and the
electron localized on the sublevel $a$, in the first approximation
with respect to $W(x)$ the probability of the backscattering per
unit time is~\cite{Dav}

\begin{equation}
P=\frac{2\pi}{\hbar}\left|\langle\Phi_{f}|W|\Phi_{in}\rangle
\right|^2 \eta(U_{f}), \label{3bar}
\end{equation}

\noindent where $\eta(U_{f})$ is the density of final states at the
energy $U_{f}$, $\langle\Phi_{f}|$ is the final wavefunction and
$|\Phi_{in}\rangle$ is the initial wavefunction combined of injected
and localized electrons.

If electrons form the singlet spin configuration ($\sigma_u=
\uparrow$, $\sigma_v=\downarrow$ or $\sigma_u=\downarrow$,
$\sigma_v=\uparrow$), then spatial parts of wavefunctions have the
symmetric combination

$$\Phi_{in}(x_1,x_2)=u(x_1)v(x_2)+u(x_2)v(x_1),$$
$$\Phi_{f}(x_1,x_2)=u(x_1)\bar v(x_2)+u(x_2)\bar v(x_1),$$

\noindent where $\bar v(x)$ is the wavefunction of the backscattered
electron described by Eq. (\ref{2bar}) with the substitution
$q\rightarrow -q$. For the singlet spin state the backscattering
probability (\ref{3bar}) is equal to

\begin{equation}
P_S=\frac{8\pi\eta(U_{f})}{\hbar}\left|A+B\right|^2, \label{4bar}
\end{equation}
\noindent where

$$A=\int_0^d u^{*}(x_1){\bar v}^{*}(x_2)W(x_1-x_2)u(x_1)v(x_2)\,
dx_1dx_2,$$
$$B=\int_0^d u^{*}(x_2){\bar v}^{*}(x_1)W(x_1-x_2)u(x_1)v(x_2)\,
dx_1dx_2,$$

If electrons form the triplet spin configuration ($\sigma_u=
\uparrow$, $\sigma_v=\uparrow$ or $\sigma_u=\downarrow$,
$\sigma_v=\downarrow$), then spatial parts of wavefunctions are
antisymmetric

$$\Phi_{in}(x_1,x_2)=u(x_1)v(x_2)-u(x_2)v(x_1),$$
$$\Phi_{f}(x_1,x_2)=u(x_1)\bar v(x_2)-u(x_2)\bar v(x_1).$$

\noindent For the triplet state the probability (\ref{3bar}) can be
written as

\begin{equation}
P_T=\frac{8\pi\eta(U_{f})}{\hbar}\left|A-B\right|^2. \label{5bar}
\end{equation}

\noindent Magnitudes $A$ and $B$ in relations (\ref{4bar}) and
(\ref{5bar}) are functions of wavevectors $q$ and $k_a$. Besides,
the wavevector $k_a$ depends on the number $n$ of localized level:
$k_ad=\pi(n+1/2)$. Taking into account wavefunction forms
(\ref{1bar}) and (\ref{2bar}), for the uniform interaction $W(x)=W$
we obtain

$$A=\frac{C_u^2C_v^2W(k_ad+1)}{4ik_a^2q^2}\left[\exp(2iqd)-1\right]$$
$$B=\frac{C_u^2C_v^2W}{2k_aq(k_a^2-q^2)^2}\left[(-1)^n\exp(iqd)(iq+k_a)\right.$$
$$\left.-(iq-k_a)\right]^2.$$

Probabilities $P_S$ (\ref{4bar}) and $P_T$ (\ref{5bar}) strongly
depend on the difference of wavevectors $\Delta q=q-k_a$ and,
consequently, on the difference $\Delta U$ between the energy of
injected electron $U_v=U+\Delta U$ and the energy of localized
electron $U$ in the QW. For $\Delta U\ll U$ the energy difference is

\begin{equation}
\Delta U=\hbar\Delta q\cdot \sqrt{\frac{2U}{m}}=
\frac{\hbar^2k_a\Delta q}{m}.\label{X1}
\end{equation}

\begin{figure}
\begin{center}
\includegraphics*[scale=.38]{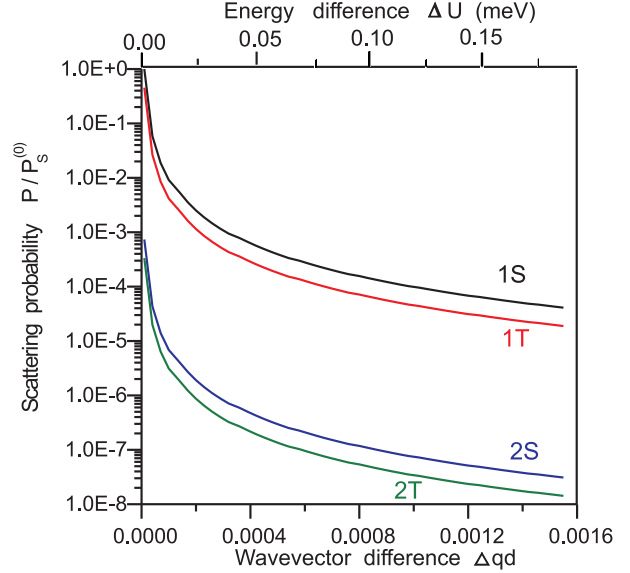}
\end{center}
\caption{Probabilities of singlet and triplet backscattering of
injected electron with the wavevector $q$ on the QW versus the
normalized wavevector difference $\Delta qd=(q-k_a)d$.
$k_a=\pi(n+1/2)/d$ is the wavevector of localized electron, $n$ is
the number of level. Probabilities are normalized by the magnitude
of the singlet probability $P_S^{(0)}$ with $n=0$ and $q=k_a$. 1S,
1T are singlet and triplet backscattering on the first level
($n=0$); 2S, 2T are singlet and triplet backscattering on the second
level ($n=1$), respectively. For the QW with the width $d=1$~nm and
the energy depth $U=94$~meV the singlet and triplet probabilities 1S
and 1T are shown as functions of the difference $\Delta U$ between
the energy of injected and localized electrons (upper axis).}
\label{bar3}
\end{figure}

\noindent Singlet and triplet backscattering probabilities versus
the normalized wavevector difference $\Delta qd$ for $n=0,1$ are
shown in Fig. \ref{bar3}. Probabilities are normalized by the
magnitude of the singlet probability $P_S^{(0)}$ with $n=0$ and
$q=k_a$. In accordance with relation (\ref{X1}), the singlet and
triplet probabilities 1S and 1T ($n=0$) are shown as functions of
the variable $\Delta U$ (upper axis). In this case the QW contains
only one exchange-splitted level. Calculations have been done for
$d=1$~nm and $U=94$~meV. It is necessary to notice that the
probability of the singlet backscattering $P_S$ (curves 1S, 2S) is
higher than the triplet backscattering probability $P_T$ (curves 1T,
2T) -- the backward scattering becomes dependent on spins of
injected electrons. For scattering of injected electrons on the
level with $n=0$ and with the wavevector $q\rightarrow k_a=\pi/2d$,
the ratio of singlet and triplet probabilities leads to the relation

$$\frac{P_S}{P_T}\rightarrow\left(\frac{\pi+3}{\pi+1}\right)^2=2.20.$$

The backscattering probability strongly reduces with growth of
$\Delta q$ and $\Delta U$. The greatest magnitude of backscattering
is achieved for the level with $n=0$. Thus, the backscattering
process becomes important, if (1) the QW contains only
exchange-splitted level with $n=0$, (2) one sublevel of the
exchange-splitted level with certain spin orientation lies at the
top of the QW and (3) the energy of injected electrons is closed to
the energy of localized electron on this sublevel.

\section{Reduction of the current}

The capture of backscattered electrons by the QW leads to the
considerable current reduction dependent on spin orientation of
injected electrons. If the Fermi level lies below localized electron
levels in the QW, then at a finite temperature these levels are
partially filled by electrons. Backscattered electrons are captured
by the QW and, in accordance with their spin orientation, they
occupy different localized levels. We suppose that the spin
relaxation time is greater than the storage time of additional
electrons in the QW. Then, for the singlet scattering process
backscattered electrons occupy the sublevel $b$ with spin
orientation opposite to spin orientation of the sublevel $a$ (Fig.
\ref{bar2}). The sublevel $b$ lies below the sublevel $a$. On the
contrary, for the triplet case backscattered electrons fall on the
sublevel $a$. The storage time $\tau$ of the presence of additional
electrons in the QW depends on the electron-hole recombination, on
temperature activation processes, and on the electron tunneling into
the conduction band. For underlying levels the storage time $\tau$
is greater than the storage time of electrons on overlying ones. The
additional charge in the QW leads to the electrostatic blockade of
injected electrons and to the current reduction. In this way, the
current flowing in ferromagnetic metal / semiconductor
heterostructure with QW, which contains exchange-splitted levels, is
unstable. This current instability is accompanied by the charge
accumulation in the QW and by the current reduction depended on spin
orientations of injected electrons.

Let us calculate the reduction of the current. For clarity, we
consider the current reduction caused by the singlet backscattering.
In the triplet case, the consideration is analogous. The current
density flowing over the QW is equal to

\begin{equation}
j=en\mu E, \label{6bar}
\end{equation}

\noindent where $e$ is the electron charge, $\mu$ is the electron
mobility, $E$ is the electric field,

$$n=n_0\exp\left(\frac{-e\varphi}{kT}\right)$$

\noindent is the electron concentration over the QW, $n_0$ is the
electron concentration without an electric field, $\varphi$ is the
potential of the field of additional localized electrons in the QW,
$k$ is the Boltzmann constant, and $T$ is the temperature. In the
singlet backscattering case, the additional charge accumulates on
the sublevel $b$ (Fig. \ref{bar2}). The potential $\varphi$ of the
field caused by this additional charge is determined by the
equation~\cite{Land}

\begin{equation}
\frac{d^2\varphi}{dx^2}=\frac{4\pi e}{\varepsilon}(n_b-n_b^{(0)}),
\label{7bar}
\end{equation}

\noindent where $\varepsilon$ is the dielectric permittivity of the
semiconductor in the QW region; $n_b$ and $n_b^{(0)}$ are electron
concentrations on the sublevel $b$ in the electric field and without
a field, respectively. If the additional concentration of the charge
$n_b-n_b^{(0)}$ is uniformly distributed over the QW width, then the
solution of Eq. (\ref{7bar}) is given by

$$\varphi(x)=\frac{2\pi e}{\varepsilon}(n_b-n_b^{(0)})x^2.$$

\noindent Injected electrons must surmount the additional barrier
with the energy height

\begin{equation}
e\varphi=\frac{2\pi e^2}{\varepsilon}(n_b-n_b^{(0)})d^2.
\label{8bar}
\end{equation}

\noindent Taking into account relations (\ref{6bar}) and
(\ref{8bar}), we obtain the current density of electrons incoming on
the sublevel $b$

$$j_b=P_Sj=P_Se\mu En_0\exp\left[\frac{-2\pi e^2(n_b-n_b^{(0)})d^2}{\varepsilon
kT}\right].$$

\noindent Release of additional electrons from the sublevel $b$ is
determined by the time $\tau_b$ and the current density of outgoing
electrons can be written as

$$\bar j_b=\frac{e(n_b-n_b^{(0)})d}{\tau_b}.$$

\noindent For the equilibrium process $j_b=\bar j_b$ and

\begin{equation}
P_S\mu En_0\exp\left[\frac{-2\pi e^2(n_b-n_b^{(0)})d^2}{\varepsilon
kT}\right]=\frac{(n_b-n_b^{(0)})d}{\tau_b}. \label{9bar}
\end{equation}

\noindent Relation (\ref{9bar}) is the equation in the unknown
additional electron concentration $n_b-n_b^{(0)}$. Taking into
account relation (\ref{6bar}), we find the current reduction caused
by the singlet electron backscattering on the QW

\begin{equation}
R_S=\frac{j}{j_R}=\exp\left[\frac{2\pi
e^2(n_b-n_b^{(0)})d^2}{\varepsilon kT}\right].\label{10bar}
\end{equation}

\begin{figure}
\begin{center}
\includegraphics*[scale=.41]{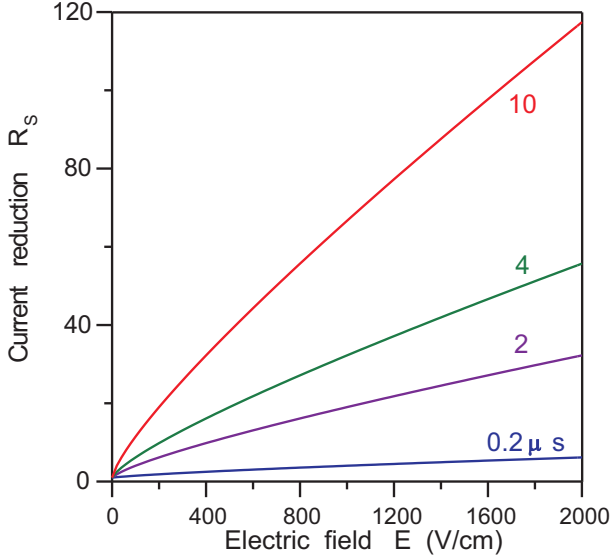}
\end{center}
\caption{Current reduction $R_S$ caused by the singlet electron
backscattering in ferromagnetic metal / insulator / semiconductor
heterostructure with quantum well (QW) contained exchange-splitted
levels versus the applied electric field $E$ for different storage
time $\tau_b$ of additional electrons in the QW. The backscattering
probability $P_S=2\cdot 10^{-6}$, width of the QW $d=$~2~nm,
temperature $T=$~300~K, and the electron concentration over the QW
$n_0=2.5\cdot 10^{17}$~cm${}^{-3}$.} \label{bar4}
\end{figure}

Current reductions $R_S$ versus the applied electric field $E$ for
different times $\tau_b$ are shown in Fig. \ref{bar4}. Calculations
are performed for $P_S=2\cdot 10^{-6}$, width of the QW $d=$~2~nm,
permittivity $\varepsilon=$~1, $T=$~300~K, $\mu=8\cdot
10^3$~cm${}^2$/V$\cdot$s, and $n_0=2.5\cdot 10^{17}$~cm${}^{-3}$.
From the presented dependencies we can see that backscattering of
injected electrons on exchange-splitted levels and accumulation of
electrons in the QW leads to considerable reduction of the current
depended on its spin orientation.

The current reduction depends on the electron concentration $n_0$ in
the semiconductor. For small values of $n_0$ the additional
concentration $n_b-n_b^{(0)}$ in Eq. (\ref{9bar}) leads to zero and
the reduction is small, $R_S\rightarrow 0$. For great values of the
concentration $n_0$ (for example, close to metal concentrations) the
QW contains filled levels and the additional charge in the QW is
impossible. As a result of this, there is no any reduction of the
current.

For the triplet backscattering case, backscattered electrons
accumulate on the level $a$. The current reduction $R_T$ is
determined by relation (\ref{10bar}), in which we must perform the
substitution $n_b-n_b^{(0)}\rightarrow n_a-n_a^{(0)}$. The
additional electron concentration $n_a-n_a^{(0)}$ is the solution of
Eq. (\ref{9bar}) with substitutions $\tau_b\rightarrow\tau_a$ and
$P_S\rightarrow P_T$. In comparison with the singlet case, for
$\tau_a \ll \tau_b$ and $P_T < P_S$ the current reduction $R_T$
caused by the triplet backscattering and by the accumulation of
electrons in the QW is insignificant.

\section{Spin polarization of electron current}

\begin{figure}
\begin{center}
\includegraphics*[scale=.53]{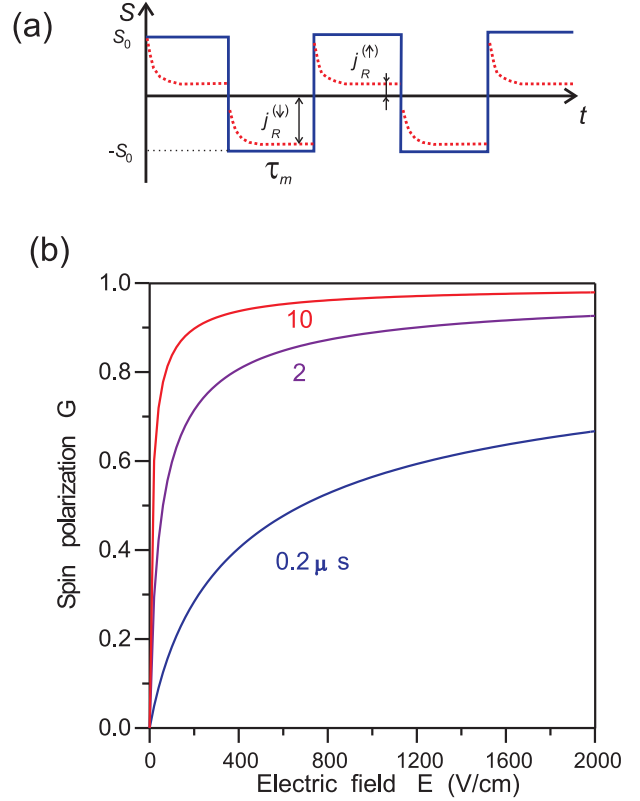}
\end{center}
\caption{(a) Current with square modulation of the spin
polarization. $j^{(\uparrow)}_R$ and $j^{(\downarrow)}_R$ are
currents with spin polarization $\uparrow$ and $\downarrow$,
respectively, after electron backward scattering on the QW. (b) Spin
polarization $G$ of the electron current caused by the electron
backscattering versus the electric field $E$ for different values of
the storage time $\tau_b$.} \label{bar5}
\end{figure}

The QW with exchange-splitted levels can be regarded as spin filter
for injected electrons. Let us consider the spin current flowing
over the QW with square modulation of the spin projection $S(t)$
(Fig. \ref{bar5}a)

\begin{equation}
j^{(\alpha)}=S(t)n\mu E,\label{11bar}
\end{equation}

\noindent where $n$ is the electron concentration, $\mu$ is the
mobility, $E$ is the electric field, $\alpha=\uparrow,\downarrow$.
We suppose that the modulation period $\tau_m$ is much greater than
the storage times $\tau_a$ and $\tau_b$: $\tau_m\gg\tau_b>\tau_a$.
Without spin-dependent backscattering and charge accumulation in the
QW the spin current $j^{(\alpha)}$ is not modified. In the presence
of singlet and triplet backscattering and accumulation of
backscattered electrons in the QW, the spin current $j^{(\alpha)}$
decreases to $j^{(\alpha)}_R$ and the reduction becomes dependent on
its spin projection $S(t)$. The magnitude of the reduction of the
current $j^{(\uparrow)}\rightarrow j^{(\uparrow)}_R$ caused by the
singlet backscattering on the sublevel $a$ (Fig. \ref{bar2}) is
higher than the magnitude of the reduction of the current
$j^{(\downarrow)}\rightarrow j^{(\downarrow)}_R$ caused by the
triplet scattering process. Taking into account relations
(\ref{6bar}), (\ref{10bar}) and (\ref{11bar}), for time regions far
from pulse edges we can write the spin polarization as

$$G=\frac{|j^{(\downarrow)}_R|-|j^{(\uparrow)}_R|}{|j^{(\downarrow)}_R|+|j^{(\uparrow)}_R|}=
\frac{R_S-R_T}{R_S+R_T}.$$

The spin polarization $G$ versus the electric field $E$ has been
calculated for different values of the storage time $\tau_b$ for
backscattering probability $P_S=2.20\cdot P_T=2\cdot 10^{-6}$, width
of the QW $d=$~2~nm, permittivity $\varepsilon=$~1, temperature
$T=$~300~K, electron mobility $\mu=8\cdot 10^3$~cm${}^2$/V$\cdot$s,
concentration $n_0=2.5\cdot 10^{17}$~cm${}^{-3}$, and time
$\tau_a=$~10~ns (Fig. \ref{bar5}b). One can notice that the spin
polarization $G$ increases with growth of the electric field $E$ and
the storage time $\tau_b$.

\section{Conclusion}

The backward scattering of injected electrons on exchange-splitted
levels of quantum wells in ferromagnetic metal / insulator /
semiconductor heterostructures can be used as the effective way of
the spin polarization of the current. The necessary condition to
obtain high values of the spin polarization is: one of the
exchange-splitted levels must be in the top region of the QW. If the
energy of injected electrons is close to the energy of localized
electrons, the backward scattering becomes dependent on spins of
injected electrons -- on singlet or triplet spin configurations. It
is found that the probability of the singlet backscattering $P_S$ is
higher than the triplet backscattering probability $P_T$. The
capture of backscattered electrons by the QW leads to an additional
Coulomb repulsion for electrons and to the considerable
spin-dependent reduction of the current flowing in the
heterostructure. The spin polarization $G$ of the current increases
with growth of the applied electric field and the storage time of
electrons in the QW and its high values can be achieved at room
temperature. In this way, the QW with exchange-splitted levels in
ferromagnetic metal / insulator / semiconductor heterostructures can
be regarded as spin filter.

\section*{Acknowledgments}
This work was supported by the Government of Russia (project No.
14.Z50.31.0021, leading scientist M. Bayer).


\end{document}